# Mass loss in metal deficient Red Giants

V. Castellani[1,2] and M. Romaniello[3]

[1] Dipartimento di Fisica, Università di Pisa, Piazza Torricelli 3, I-56126 Pisa, Italy
[2] Osservatorio Astronomico Collurania, I-64100 Teramo, Italy
[3] Scuola Normale Superiore, Piazza dei Cavalieri 7, I-56126 Pisa, Italy; e-mail: martino@cibs.sns.it

**Abstract.** We investigate the final fate of metal deficient (MD) H burning Red Giants under various assumptions about the amount of mass loss. Adopting a metallicity $Z= 10^{-8}$, we follow the evolution of 0.8 $M_\odot$ models from the Main Sequence phase till the onset of the He flash or the final cooling as a He White Dwarf, according to the amount of mass loss. We show that MD Red Giants require peculiarly large H-rich envelopes to support the H-shell burning and to allow the final ignition of He, thus favouring the occurrence of He degenerate dwarfs either as the result of single star evolution or in MD binary systems. The evolution of our models along the He burning phase or as cooling White Dwarf is briefly discussed.

## 1. Introduction

In a previous paper (Castellani, Luridiana & Romaniello 1994: Paper I), we studied the properties of so called "Red Giant Stragglers" (RGS), as produced by Red Giant stars missing the onset of the He flash for loosing too much of their H-rich envelope during the ascent of the RG branch. In that paper, we found that the critical mass of envelope supporting a RG structure increases when the star luminosity and/or metallicity decreases. According to the discussion presented in that paper, this behavior can be easily interpreted in terms of the effects of both luminosity and metallicity on the width of the H-shell burning region, larger widths requiring larger envelopes.

According to such a scenario, it appears obviously interesting to move the investigation toward the limit of metal deficient giants, where CNO burning is depressed in favor of pp reactions, much less sensitive to temperature variation. As a consequence, one expects in MD giants a maximum width of the H shell burning and, in turn, a maximum value for the critical envelope as defined above. Accordingly, MD giants should be, at least in principle, the most "risky" RG structures, i.e., the RG structures which more easily can escape the red giant branch to become degenerate He White Dwarfs. Whether or not such an occurrence can affect the evolution of single MD stars is dependent on the assumptions about mass loss from similar stars. However, this implies, again at least in principle, that MD binary systems should be more efficient in producing He degenerate WD.s, an occurrence which could play a role in the evolutionary scenario concerning the early history of the Galaxy.

This note is devoted to approach the above referred problem, presenting a suitable set of evolutionary computations to give light on the evolutionary fate of MD Red Giants. In the next section we will present theoretical results concerning the evolution of selected models along the red giant branch, discussing the main features of the straggling giants. In the final section, the evolution of both He burning models and cooling He white dwarfs will be briefly discussed.

## 2. Metal deficient giants

Recent explorations of the evolutionary behavior of metal deficient stars (Cassisi & Castellani 1993) covering the metallicities log Z=-6 and -10 already revealed that even an amount of metals this small can play a relevant role in the evolutionary history of low mass stars. As a matter of fact, one finds that metal deficient red giants with log Z=-6 still succeed in using the few CNO nuclei to allow H-shell CNO burning to partially support the structure, dumping the increase of internal temperatures. On the contrary, if log Z=-10 the stars are supported by pp burning only, and the internal temperatures increase till succeeding to produce fresh C which allowes a renewed efficiency of the CNO cycle.

Correspondingly, one finds that the log Z=-6 models follow well known theoretical prescriptions concerning me-amount of heavy elements down to log Z=-6 the mass of the He core at the He flash keeps increasing, whereas the luminosity of the tip of the giant branch continuously decreases. However, when going down to log Z=-10 this behavior is partially reversed, with the tip luminosity still decreasing but with the mass of the He core coming back to smaller values, an occurence recently confirmed by Fujimoto et al (1995). Taking into account similar evidences, in investigating metal deficient giants we adopted log Z=-8, adding to the main goal of investigating MD RG stragglers the interesting and more general investigation of the behavior of stellar structures in such an unexplored value of metallicity. Correspondingly, we will assume an original amount of Helium as given by Y=0.23, i.e., a value which should be close to the primeval production of He and, thus, adequate for the investigated early generation of stars.

As in Paper I, computations have been performed assuming in all cases an original mass of 0.8 $M_\odot$ (which is a typical value for the evolving mass in Globular clusters, corresponding to an age of about 14 Gyr for Population II), but for various assumptions about the value of the parameter $\eta$ governing the efficiency of mass loss, as defined in the same Paper I. However, the results of a similar investigation can be regarded as reasonably independent of the assumed original mass, since the evolution of low mass Red Giants is largely independent of such a parameter. Our models take into account all relevant thermodynamical properties of stellar matter (Straniero 1988), including electrostatical interactions between ions (Ogata & Ichimaru 1987) and electron degeneracy. Interior opacies are taken from the Los Alamos Opacity Library (LAOL, Huebner et al 1977), which have been shown to be fully adequate for the low range of metallicities (Cassisi & Castellani 1993, but see also Cassisi et al 1995). The models have been followed from their initial main sequence location up to the onset of the He flash or, if escaping the RGB, all along their WD phase until reaching a vanishing contribution of pp burning, i.e., down to log $L/L_\odot \simeq$ -4.5. In spite of the very low value of the metallicity, one finds that our giants succeed in being progressively powered by CNO reactions, which can contribute up to about 95% of the total luminosity in structures approaching the He flash.

Table 1 compares selected physical quantities for our model without mass loss with similar models with log Z=-4, -5, -6 and -10 computed by Cassisi & Castellani (1993) and Cassisi et al (1995).

One finds that, decreasing the metallicity, the top luminosity of the giant branch continuously decreases. This is not the case for the mass of the He core in flashing giants, which reaches a maximum when log Z=-8, thus decreasing when going down to log Z=-10. Note that the contrasting behavior of luminosity and He cores with Z obviously indicates that the metallicity affects the core mass-luminosity relation of flashing giants. As a matter of fact, for each given value of the He core mass $M_c$ one finds that the luminosity of the corresponding giant becomes sensitively larger when metallicity is increased above log Z=-10 (see, e.g., data in Table 3 in Cassisi & Castellani 1993). As a result, one finds that from log Z=-4 down to log Z=-8 low mass RG.s behave like a "normal" tough extreme Pop.II stars, where a decrease in metallicity requires larger He cores to ignite He burning. Only at the extreme case log Z=-10, when the inner temperatures grow to cope with the lack of CNO burning till producing fresh C nuclei, the heating of the stars begins to dominate, thus allowing the ignition of He with smaller He cores and, consequently, at even smaller luminosities. One could thus speculate of the opportunity of reserving the name of population III stars only to structures with original Z below log Z=-8, showing the behavior of our log Z=-10 model.

**Table 1.** Selected parameters for models without mass loss

| log Z | $M_{c,flash}$ | log $L_{flash}$ |
|---|---|---|
| -4  | 0.504 | 3.248 |
| -5  | 0.516 | 3.156 |
| -6  | 0.529 | 3.055 |
| -8  | 0.537 | 2.722 |
| -10 | 0.507 | 2.376 |

Our results for the dependence of the core mass at the flash with the original metal content appear in excellent agreement with those presented by Fujimoto et al (1995, their Table 1). There seems to be only a little systematic difference between our and their values (i.e. their masses are 0.003 $M_\odot$ greater than ours for small metallicities), probably due to a slightly different definition of core mass, either at the point where the shell burning reaches it maximum, as we do, or at the point where the hydrogen abundance is half of the surface value, as they do.

Coming to the main target of the present investigation, Figure 1 gives the evolutionary paths in the HR diagram of 0.8 $M_\odot$ H burning models for selected assumptions about the efficiency of mass loss, whereas Table 2 lists selected physical quantities depicting the behavior of internal structures for the labeled values of $\eta$. Left to right one finds: the value of $\eta$, the final fate of the giants, the mass $M_c$ of the He core, the total mass $M_{tot}$, the mass of the envelope $M_e$ and the luminosity of the star either at the flash ($\eta$=0, 1) or when escaping the RG branch and crossing log $T_e$= 3.80 and, finally, the mass of the He core when the giant reaches log $L/L_\odot$=1.5 along the RGB.

A quick inspection of the data shows that, as expected, mass loss has a negligible effect on the mass of the He core at the flash and, more generally, on the core mass-luminosity relation. Again as expected, the comparison with data in Paper I given in Figure 2 shows that for each given value of $\eta$ the amount of mass loss in a flashing giants decreases with the metallicity.

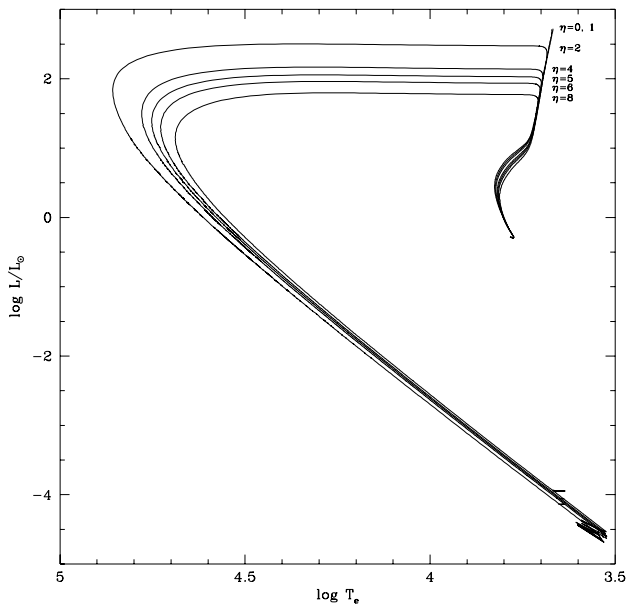

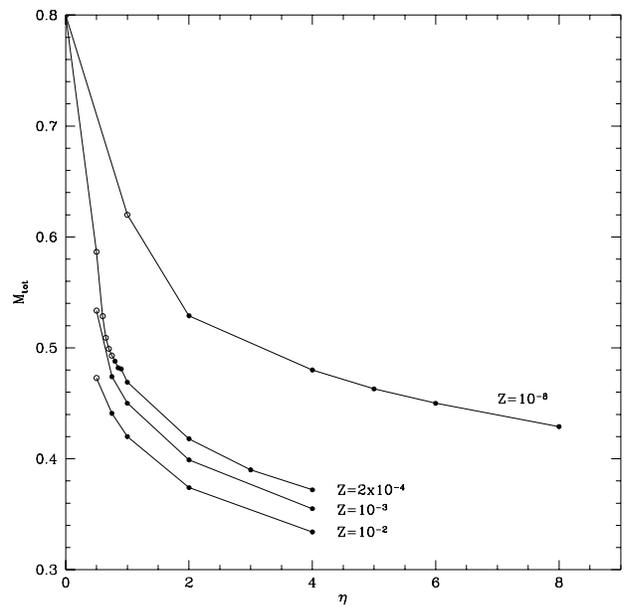

**Fig. 1.** The evolutionary paths in the HR diagram of 0.8 $M_\odot$ models with Y=0.23, log Z=-8 for the various labeled assumptions of the mass loss parameter $\eta$

**Fig. 2.** The final mass of Red Giants at the flash (open circles) as a function of $\eta$ for the labeled assumptions on the star metallicity. Filled circles give the total mass at log L/$L_\odot$ for stars escaping the RG branch to cool down as He WD.s.

**Table 2.** Selected parameters for models with mass loss

| $\eta$ | RGF   | $M_c$ | $M_{tot}$ | $M_e$ | log L | $M_c$(log L=1.5) |
|--------|-------|-------|-----------|-------|-------|------------------|
| 0      | He Fl | 0.537 | 0.800     | 0.263 | 2.72  | 0.32             |
| 1      | "     | 0.536 | 0.620     | 0.084 | 2.71  | 0.32             |
| 2      | Escap.| 0.492 | 0.529     | 0.036 | 2.45  | 0.32             |
| 4      | "     | 0.434 | 0.480     | 0.046 | 2.14  | 0.32             |
| 5      | "     | 0.415 | 0.463     | 0.049 | 2.03  | 0.33             |
| 6      | "     | 0.398 | 0.450     | 0.052 | 1.94  | 0.33             |
| 8      | "     | 0.373 | 0.429     | 0.056 | 1.77  | 0.33             |

However, one can attach to this occurrence only little significance, since it depends on the particular formulation for mass loss rates (the one suggested by Reimers) we are relying on. As a matter of fact, our approach has to be regarded only as a reasonable way to modulate the amount of mass loss, without attaching too much significance to the precise dependence on stellar parameters, and in particular on the radius and, thus, on the metallicity of the giant.

Much more interestingly, Fig. 3 reveals the relevant increase of the critical envelopes in log Z=-8 stars, i.e., it confirms that metal deficient giants need a much more massive H-rich envelope to keep on with their evolution.

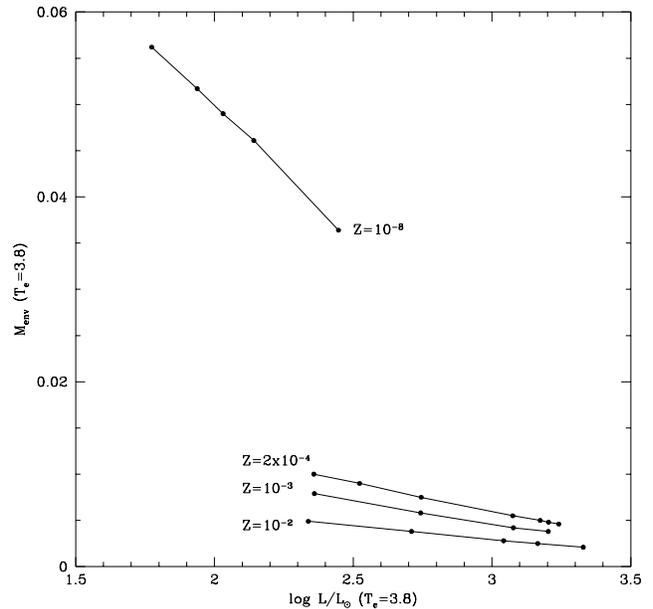

**Fig. 3.** The mass of the envelope for stars leaving the giant branch when crossing log $T_e$=3.8 as a function of the corresponding luminosity and for the labeled assumptions about the star metallicity.

giants filling their Roche's lobe in a binary system should easier turn off evolution toward degenerate structures producing, for each given luminosity, most massive dwarfs for leaving the RG branch with both more massive He cores and more massive H-rich envelopes.

## 3. Post-giant evolution

To complete the theoretical scenario, let us here briefly discuss the evolutionary fate of the giants presented in the previous section. Red Giants succeeding in igniting He have a well defined mass of the He core, and thus distribute along the corresponding ZAHB locus, according to the amount of mass loss. Fig. 4 shows the evolution during central He burning and beyond of ZAHB models with selected values of the stellar mass, if the "internal pollution" suggested by Fujimoto et al (1995) is neglected.

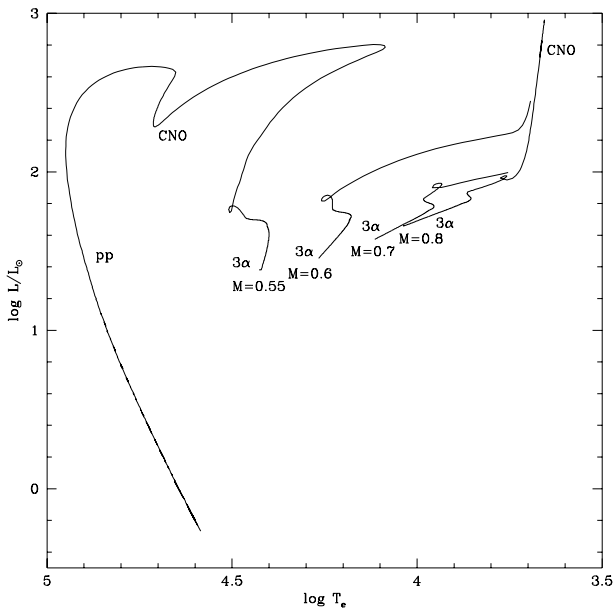

**Fig. 4.** The evolution off ZAHB of selected models for log Z=-8 with the labeled values of total masses and for the common value of the mass of the ZAHB He core appropriate for this metallicity, i.e. 0.537 $M_\odot$. Details of the prevailing energy sources are labeled along the tracks.

According to the increased mass of the He core, we now know that for log Z=-8 the ZAHB distribution is the most luminous ZAHB expected from old stellar populations. Details of the structural evolution are summarized in the same figure, not deserving further comments. We only notice that the transition between central and shell He burning is supported by the partial contribution of pp burning only, whereas a substantial contribution of self burning phase, where we stopped our computations. The lack of CNO nuclei allows the 0.55 $M_\odot$ to become a cooling white dwarf after the exhaustion of He shell burning passing through a first phase of self produced CNO burning and a final stage of pp burning before the final thermal cooling.

As for the evolution of RG stragglers, when the star leaves the RG branch it is mainly supported by CNO burning, which remains the main source of energy while the star evolves quietly to ingress its cooling sequence down to about log L/L$_\odot \simeq$1.5, where the pp begins to dominate until nuclear burning is eventually switched off at about log L/L$_\odot \simeq$-4.5. In Table 3 we compare the mass of the H rich envelope in two selected points of the off-RG evolution, namely when log L/L$_\odot$=1 and log L/L$_\odot$=-4.5, with the mass of the envelope when the star was crossing log T$_e$=3.8.

**Table 3.** Selected parameters for models along the cooling sequence

| $\eta$ | $M_e$(3.8) | $M_e$(L=1) | $M_{e,fin}$ | $M_{c,fin}$ |
|---|---|---|---|---|
| 2 | 0.036 | 0.0021 | 0.00012 | 0.52255 |
| 4 | 0.046 | 0.0029 | 0.00016 | 0.46942 |
| 5 | 0.049 | 0.0031 | 0.00017 | 0.45158 |
| 6 | 0.052 | 0.0034 | 0.00019 | 0.43668 |
| 8 | 0.056 | 0.0043 | 0.00019 | 0.41297 |

From these data, it appears that the large majority of the envelope is burned into He at luminosities larger than log L/L$_\odot$=1, where one expects a slowing down of the cooling rates. However, the evolutionary times in this phase are so short, i.e. of the order of 5x10$^6$ years from the RGB tip down to log L/L$_\odot$=1 on the cooling sequence, that one expects a negligible evidence of hot WD.s, unless unexpectedly abundant sample of RGS are present in old stellar populations.

In any case, even the small amount of envelope left at log L/L$_\odot \leq$ 1 can significantly influence the evolution at low luminosities. This is shown in Fig. 5, where we report the evolution with time of the stellar luminosity along the cooling sequence. As a matter of fact, the model with $\eta$=8, wich enters the cooling sequence with an envelope twice as large than the model with $\eta$=2 (see Table 3), is significantly slower than the other models. This is due to the residual pp burning whose efficiency increases with the envelope mass. The spikes that appear in the same figure are caused by weak pp shell flashes whose strenght and extent to lower luminosities increase as the degeneracy in the active shell, i.e. the total mass, increases.

As a conclusion, decreasing Z down to log Z=-8, one finds stars with a maximum size of the He core at the

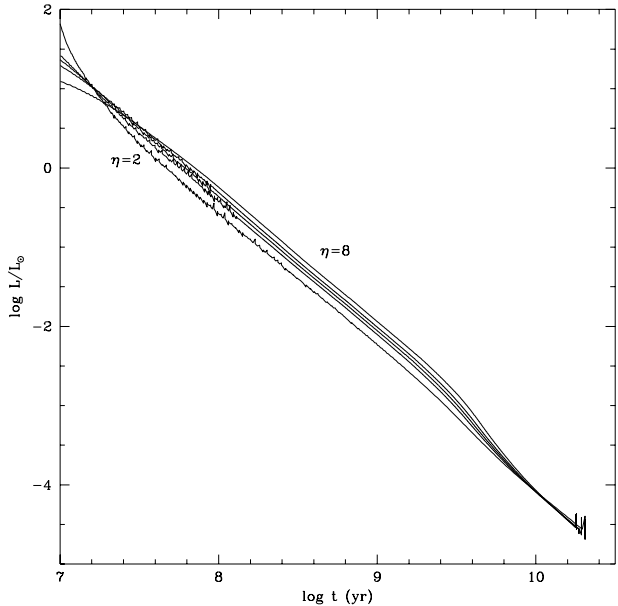

**Fig. 5.** The evolution with time of the luminosity of RG Stragglers along their cooling sequence. Times are normalized to $10^7$ years at the hottest point of the evolution.

He flash or maximum size of the H rich envelope of RG Stragglers leaving the RG branch for excess of mass loss. For these last structures, the lack of CNO nuclei allows a quiet H burning during the cooling sequence, avoiding the shell flashes which put rather severe constraints on the evolution of more metallic He White Dwarfs (Paper I). As whole, a set of characteristic features marking the evolution these extreme representants of Poulation II evolution.

*Acknowledgements.* We thank Santi Cassisi for providing us with his modified version of the FRANEC evolutionary code suitable for low metallicity evolutions, as well as for useful discussions on the subject. We would like also to thank the referee, M. Forestini, for his useful suggestions.